\documentclass[aps,prb,groupedaddress,twocolumn]{revtex4-1}
\pdfoutput=1

\usepackage{amsmath}
\usepackage{amssymb}
\usepackage{graphicx}
\usepackage{bm}
\usepackage{hyperref}
\hypersetup{%
breaklinks = {true},
citecolor = {blue},
colorlinks = {true},
linkcolor = {red},
pdfauthor = {\textcopyright\ Johan Nilsson},
pdfcreator = {\LaTeX\ and \flqq hyperref\frqq},
pdffitwindow = {true},
pdfmenubar = {true},
pdfpagelayout = {SinglePage},
pdfstartview = {Fit},
pdftoolbar = {true},
plainpages = {false},
}

\usepackage{bbm}

\newcommand{\la}{\langle}
\newcommand{\ra}{\rangle}

\newcommand{\dla}{\langle \negthinspace \langle}
\newcommand{\dra}{\rangle \negthinspace \rangle}

\newcommand{\ua}{\uparrow}
\newcommand{\da}{\downarrow}

\newcommand{\vk}{{\bf k}}

\usepackage{bbold}
\newcommand{\eye}{\mathbb{1}}

\usepackage{color}
\usepackage{color}

\begin{document}


\title{The Hubbard dimer within the Green's function equation of motion approach}

\author{Francesco Catalano}
\affiliation{Department of Physics and Astronomy, Uppsala University,
P.O. Box 516, SE-75120, Uppsala, Sweden}
\author{Johan Nilsson}
\affiliation{Department of Physics and Astronomy, Uppsala University,
P.O. Box 516, SE-75120, Uppsala, Sweden}

\date{Dec 4, 2019}

\begin{abstract}

We consider a formulation of the equation of motion technique for Green's functions in which the unknown averages are computed by solving a linear system. This linear system appears formally solvable for all finite temperatures, but depending on the system parameters the condition number can be very large, making the solution numerically unfeasible at low temperatures. In the example that we consider, the Hubbard dimer, we can get rid of this problem by making use of total spin as a good quantum number.

\end{abstract}

\maketitle

\section{Introduction}

Despite a great amount of work, interacting many-body systems remain poorly understood in the most interesting regime of moderate to strong interactions. The main reason for this is the lack of a simple reference system that may be used to characterize and understand the physics of this class of systems (also more than one such reference system is needed). When such a reference system is available the physics of realistic many-body systems in the corresponding class could be obtained perturbatively starting from the reference system and would then be deemed to be ``understood''.
The workhorse of solid state physics, independent fermions, provides a versatile such a reference system. For systems outside of the independent fermion class, however, very few examples exists, although dynamical mean field theory provides a popular reference system that can account for local quantum fluctuations.\cite{DMFT_RMP}

Equations of motion (EOM) methods operates differently, in that only dynamics and matrix elements of certain types of excitations are considered. In particular EOM combined with Green's functions methods is a well-known technique of great generality, for reviews see e.g. Refs.~\onlinecite{Zubarev,Mancini_Avella_1,0038-5670-34-11-R02,RevModPhys.40.153}. Since the EOM almost never close (except in some very special models) the formalism has to be supplied with a scheme for truncating the EOM and to include the neglected operators in some approximate way. Also here very general theoretical constructions may be introduced, in particular the irreducible self-energy formalism of Tserkovnikov provides a beautiful general way to do this.\cite{Tserkovnikov1981} Although very nice on a formal level this formalism is almost too general to be useful since all of the quantities involved can typically not be evaluated from within the formalism itself. To proceed uncontrolled approximations are therefore often introduced at this stage.

Uncontrolled approximations (for example arbitrary truncations of the EOM) may also easily lead to unphysical results, for example negative spectral weights and violation of certain sum rules or constraints. By demanding certain analytic properties of the Green's function one is at least guaranteed to generate a physical approximation with positive spectral weight. Examples of formalisms that leads to a physical approximate physical theories include the Roth procedure,\cite{PhysRevLett.20.1431,PhysRev.184.451} and the irreducible self-energies of Tserkovnikov\cite{Tserkovnikov1981} (the Roth procedure may be viewed as a particular way of truncating the irreducible self-energy in this more general scheme). The problem with this and related approaches is again that it is usually not possible to calculate the necessary averages from within the theory itself without making some uncontrolled approximations. 

In this work we consider the EOM from within the Green's function formalism in a somewhat unconventional way.
In general there are two types of quantities that must be fixed to determine the Green's function for excitations with some chosen set of quantum numbers.
The first is the spectrum of the theory and the second is the spectral weight distribution.
The spectrum is the set of the positions of the poles of the Green’s function
 (or by summing up infinitely many poles with proper weights one may get branch cuts). When solving quantum mechanical problems in the grand canonical formalism exactly the spectrum is clearly independent of all of the averages in the system, so this is a property that we would like to retain in our formalism. In conventional mean field theories this is no longer the case, which may be important to capture some physical effects in a simple way, but on a microscopic level mean field theory is clearly not always correct.
A simple illustration of this is provided by the Hubbard model in the strong coupling limit where the states may (at zeroth order) be grouped according to the number ($m$) of doubly occupied sites they contain. This leads to an approximate energy spectrum $E_{m\alpha} = m U+ \epsilon_\alpha$ where $\alpha$ label the different states in each sector. Perturbatively $\epsilon_\alpha = a_{0\alpha} t + a_{1\alpha} t^2/U + \ldots$.\cite{Eskes} Here $U$ is the energy cost of double occupancy, $t$ the strength of the hopping, $a_{i\alpha}$ are some dimensionless coefficients. This implies that the true excitation energies are centered around integer multiples of $U$ for $t \ll U$. In a paramagnetic mean field theory for fermions the Hartree excitation energies are $E_{HF \alpha} = U \bar{n}+ \epsilon_{HF\alpha}$ where $2\bar{n}$ is the average number of fermions per site. In this case, it is clear that the actual excitation energies are not well captured by one static mean field $\bar{n}$.
The correct resolution of this is to include two types of excitations with energies near $0$ and $U$. With proper weights of these the average excitation energy will be $U \bar{n}$ as it should.\cite{Eskes,PhysRevB.39.6962} In this case at least two peaks are necessary in the spectral function. In the EOM language this implies that at least two operators must be included in the EOM. This is what Hubbard did by the introduction of the Hubbard operators.\cite{Hubbard_I}
In any case it is often relatively easy to know either exact or approximate energies of many excitations.

The spectral weights that must be determined are some particular matrix elements that are encoded in the Green's function. This involves averages of certain operators in the ground state (or a thermodynamical average at finite temperatures). In particular this will fix the weights of the poles of different excitations. It is usually more difficult to determine these averages than the excitation energies. In particular the averages that enters are typically not independent of each other, but the constraints and relations between different averages are very difficult to discern. There are also constraints related to the Pauli principle and additional sum rules that the exact theory should satisfy. This has been stressed and used by Mancini and co-workers to determine certain averages.\cite{Mancini_Avella_1}

In this work we consider another way to determine the necessary averages using what we to the best of our knowledge believe is a novel scheme, in which the averages are determined by simply solving a linear system. This is quite appealing from a formal point of view. For computational purposes it works very well when the linear system is non-singular. When the system is singular (which we find it to be in certain parameter regimes in the example we consider) we find that it is regularized by a finite temperature. We believe that any finite temperature actually makes the system non-singular, but for low temperatures the condition number might be so high that it is not technically feasible to directly implement this method numerically using calculations with standard double precision. We interpret the reason for this problem to be that the formalism does not know what state to do the average on. This problem may be fixed by using either continuity arguments or by fixing some conserved quantity such as the total spin.

The paper is organized as follows: In Sec.~\ref{sec:formalism} we provide a general discussion of and introduction to the formalism. In Sec.~\ref{sec:example} we solve the Hubbard dimer using our method. Here we also discuss some subtleties of the method that we just mentioned. Finally in Sec.~\ref{sec:conclusions} we provide some conclusions and an outlook.

\section{Formalism}
\label{sec:formalism}

From now on we will mainly use the notation of Tserkovnikov,\cite{Tserkovnikov1981} which is similar to that of Zubarev,\cite{Zubarev} Hubbard,\cite{Hubbard_I} and Roth.\cite{PhysRev.184.451}
Let us assume that we have a complete set of fermionic operator $\{\hat{A}_i\}_{i=1}^M$ that are closed under commutation with the Hamiltonian $\hat{H}$
\begin{equation}
\bigl[\hat{A}_{i},\hat{H} \bigr] =K_{ij} \hat{A}_j,
\end{equation}
for some matrix $K$.
Here and in the following we will use summation convention unless specified otherwise.
Let us further assume we have a fermionic operator $\hat{B}$, such that we may write
\begin{equation}
\label{av:anti}
\langle A_i | B^\dagger \rangle \equiv
\langle \{ \hat{A}_i, \hat{B}^\dagger \} \rangle = v_i + M_{ij} \langle \hat{B}^\dagger \hat{A}_j \rangle,
\end{equation}
possibly after making use of some symmetries that are obeyed by the averages. Here $M$ is a matrix and $v$ a vector, both of which are constants, i.e., they are independent of the averages. These two quantities are fixed by the algebra of the operators alone.
Exploiting the two previous equations and the EOM for the Green's function\cite{PhysRev.184.451}
\begin{equation} 
\label{EOM:1}
z \dla  A_i | B^\dagger \dra_z = \langle A_i | B^\dagger \rangle   +K_{ij} \dla A_j | B^\dagger  \dra_z ,
\end{equation}
it is possible to characterize the correlation function self-consistently.
In order to see this let us analyze the EOM for the eigenoperators of the theory $\{\hat{\psi}_i \}$, defined by 
\begin{eqnarray}
\bigl[ \hat{\psi}_{j}, \hat{H} \bigr] &=& \epsilon_j \hat{\psi}_{j} 
\qquad \text{(no sum)}
.
\end{eqnarray}
Expanding the eigenoperators in the basis operators $\hat{\psi}_{i} = L_{ij} \hat{A}_j$ the EOM becomes
\begin{eqnarray}
L_{ij} K_{jk} = \epsilon_i L_{ik},
\end{eqnarray}
which is a conventional left eigenvalue problem.
For this class of operators the Green's function assumes a very simple form
\begin{equation}
\label{gf:eig}
\dla \psi_i | B^\dagger \dra_z=\frac{1}{z-\epsilon_i} \la \psi_i | B^\dagger \ra.
\end{equation}
Consequently the Green's function for a general operator becomes
\begin{equation}
\dla A_i | B^\dagger \dra_z = \bigl[(z\eye - K)^{-1}\bigr]_{ij} \la A_j | B^\dagger \ra .
\end{equation}
In this form it is clear that $K$ has to posses real eigenvalues, despite that $K$ is not necessarily symmetric or even diagonalizable. Assuming $K$ to be diagonalizable we can find a complete set of eigenoperators of the theory $\{\hat{\psi}_i \}$ and the transformation matrix $L$ and its inverse $R = L^{-1}$. Let us also collect the energy eigenvalues into a diagonal matrix $\Lambda$ with diagonal elements (no sum) $\Lambda_{ii} = \epsilon_i$.
Using the previously defined transformation and Eq.~\eqref{gf:eig} we can write the Green's function using compact matrix notation as
\begin{equation}
\label{gf:form}
\dla A | B^\dagger \dra_z= R\frac{1}{z\eye -\Lambda} L \la A | B^\dagger \ra .
\end{equation}
Integrating this along a contour in the complex plane that encloses all of the singularities of the Green's function we can evaluate averages using\cite{PhysRev.184.451}
\begin{equation}
\langle \hat{B}^\dagger \hat{A} \rangle = \frac{1}{2\pi i}\oint dz f(z)  \dla A | B^\dagger \dra_z,
\end{equation}
here $f(\epsilon)$ is the usual Fermi function, either at finite or zero temperature.
The contour encircles the real axis.\cite{PhysRev.184.451}
Using this we obtain 
\begin{equation}
\label{eq:avinteg}
\langle \hat{B}^\dagger \hat{A} \rangle=R f(\Lambda) L \la A | B^\dagger \ra,
\end{equation}
where $f(\Lambda)$ is the diagonal matrix formed by the Fermi distributed eigenvalues of the diagonalized $K$ matrix, namely $f(\Lambda)_{ij} = \delta_{ik} f(\epsilon_k)\delta_{kj}$.
Using \eqref{av:anti} and \eqref{eq:avinteg} we can self-consistently determine the averages by solving a simple linear system
\begin{equation}
\label{eq:linearsys}
(\eye- R f(\Lambda) L M)\langle \hat{B}^\dagger \hat{A} \rangle= R f(\Lambda) L v.
\end{equation}
This is the most important equation for our proposed procedure.
This method to self-determine the Green's functions has two main advantages: the diagonalization of the matrix $K$ must be performed only once to get the averages and the correlation functions for all chemical potentials and temperatures. Furthermore the averages are  uniquely determined if the linear system is well-behaved.

There is no guarantee however that  the linear system in  \eqref{eq:linearsys} is numerically well-behaved, consequently in this method it is crucial to always check the conditioning of the linear system. The formalism may also straightforwardly be formulated with more than one operator $\hat{B}$, see below in Eq.~\eqref{eq:twoops} for an example. There is also the additional algebra involved to determine the matrix $M$ and there is also a choice involved of how to pick the operator sets $\{\hat{A}_i\}$ and $\{\hat{B}_j\}$. These issues are most easily explained by going through a particular example, which we turn to next.

\section{Hubbard dimer example}
\label{sec:example}

As an illustration of the method on a non-trivial system we will now consider the repulsive Hubbard dimer.
This has been previously studied in detail using the related composite operator method.\cite{Avella2003}
Although this is a small system where it is straightforward to directly diagonalize the Hamiltonian it will serve us well to illustrate how the method works in practice and some of the subtleties inherent in this scheme.
The Hubbard Hamiltonian is 
\begin{equation}
H=\sum_{\la i,j \ra,\sigma} t_{ij}(c^{\dagger}_{i \sigma}c_{j \sigma} + \text{h.c.}) + U \sum_i n_{i\uparrow}n_{i\downarrow}-\mu\sum_{i\sigma} n_{i\sigma}.
\end{equation}
We consider the usual case where $t_{ij} = -t < 0$ for nearest neighbor bonds and $t_{ij} = 0$ otherwise. $\mu$ is the chemical potential and $U$ the on-site interaction strength. In the numerical examples below we will measure all energies in units of $t$ which amounts to setting $t=1$.

To investigate the conventional fermion Green's function it is useful to consider fermionic operators that change the charge by $-e$ and spin by $+\hbar/2$.
Due to the finite size of the system we expect that the averages posses all of the symmetries of the system. This implies that it is sufficient to consider spin-up excitations for example. This also implies that we will use operators in the EOM that have the same quantum numbers as $c_{i\ua}^{\,}$.

\subsection{Fermionic basis operators}

Starting the EOM from $c_{1\ua}^{\,}$ and $c_{2\ua}^{\,}$ the EOM closes by including 6 additional operators. Picking $\hat{B} = c_{1\ua}^{\,}$, for example, and using symmetries we then find that it is not possible to write down the relations in Eq.~\eqref{av:anti} between the normalization matrix elements and the averages involving $\la \hat{B}^\dagger \hat{A}_i \ra$. To solve this problem we have used two different routes: \\
(I) We include additional operators in the set $\{\hat{A}_i\}$. Indeed, by using the 22 operators that are listed in App.~\ref{app:A}, we can show that $\{\hat{A}_i\}$ is closed under commutation with the Hamiltonian, and we can also find the matrix $M$ and the vector $v$. The non-zero matrix elements of  $K$, $M$, and $v$ are reported respectively in App.~\ref{app:B}. This is a brute-force way to solve the problem, and it is possible to use fewer than 22 operators. The algebra is facilitated by using computer algebra, such as for example SNEG.\cite{sneg} \\
(II) Another possibility is to use more than one operator $\hat{B}$. We find that using $\hat{B}_1 = c_{1\ua}^{\,}$ and $\hat{B}_2 = c_{2\ua}^{\,} c_{1\da}^{\dagger} c_{2\da}^{\,}$ we may write down the following relations
\begin{subequations}
\label{eq:twoops}
\begin{eqnarray}
\langle A_i | B_1^\dagger \rangle &=& v^{1}_i + M^{11}_{ij} \langle \hat{B}_1^\dagger \hat{A}_j \rangle + M^{12}_{ij} \langle \hat{B}_2^\dagger \hat{A}_j \rangle, \\
\langle A_i | B_2^\dagger \rangle &=& v^{2}_i + M^{21}_{ij} \langle \hat{B}_1^\dagger \hat{A}_j \rangle + M^{22}_{ij} \langle \hat{B}_2^\dagger \hat{A}_j \rangle.
\end{eqnarray}
\end{subequations}
Combining these with Eq.~\eqref{eq:avinteg} we arrive at a linear system to solve for the averages that are necessary to determine the Green's function. The main advantage of this scheme is that it is sufficient to consider the smaller number of basis operators, and therefore the diagonalization of $K$ is more efficient. A drawback is that there is some additional algebra involved in finding the matrices $M$ (these are also not unique since average can often be expressed in different ways in terms of composite operators), also the size of the linear system to solve is doubled in this case. We have implemented this scheme, but it does not seem to give any improvement over the other method (I) when it comes to the main issue of the condition number discussed below.

\subsection{Momentum basis}

The Hubbard dimer may also be viewed as a translational invariant system. Then all excitation may be labeled by momentum as well (in the dimer this also correspond to spatial inversion). Of course there are only two possible momenta in the dimer ($\vk = 0$ and $\vk = \pi$) but this generalizes to larger systems as well.
 In the dimer it is then found that the EOM may be written in terms of 4 operators in each momentum sector, so as usual using symmetries reduces the matrix dimensions.
This solution was used in Ref.~\onlinecite{PhysRevB.61.12816} to generate approximations in the full lattice Hubbard model, see also Ref.~\onlinecite{Avella2003}. Again it is not possible to write down the relations \eqref{av:anti} using only these 8 operators and $\hat{B}_\vk=c_{\vk \ua}$, but it is possible by extending the operator basis to include in total 16 operators.

\subsection{The spectrum of the excitations}

The $K$ matrix given in App.~\ref{app:A} is diagonalizable and have real eigenvalues for all the interaction strengths considered in this paper $U\in[1,10]$.
From Eq.~\eqref{gf:form} we see that the eigenvalues of the matrix $K$ corresponds to excitation energies of the system and they can readily be compared with the excitation energies obtained from exact diagonalization in this simple system. 
The predicted excitations energies are found to agree with the exact diagonalization result as they should.

\subsection{Condition number -- High temperatures}

\begin{figure}
\includegraphics[width=0.5\textwidth]{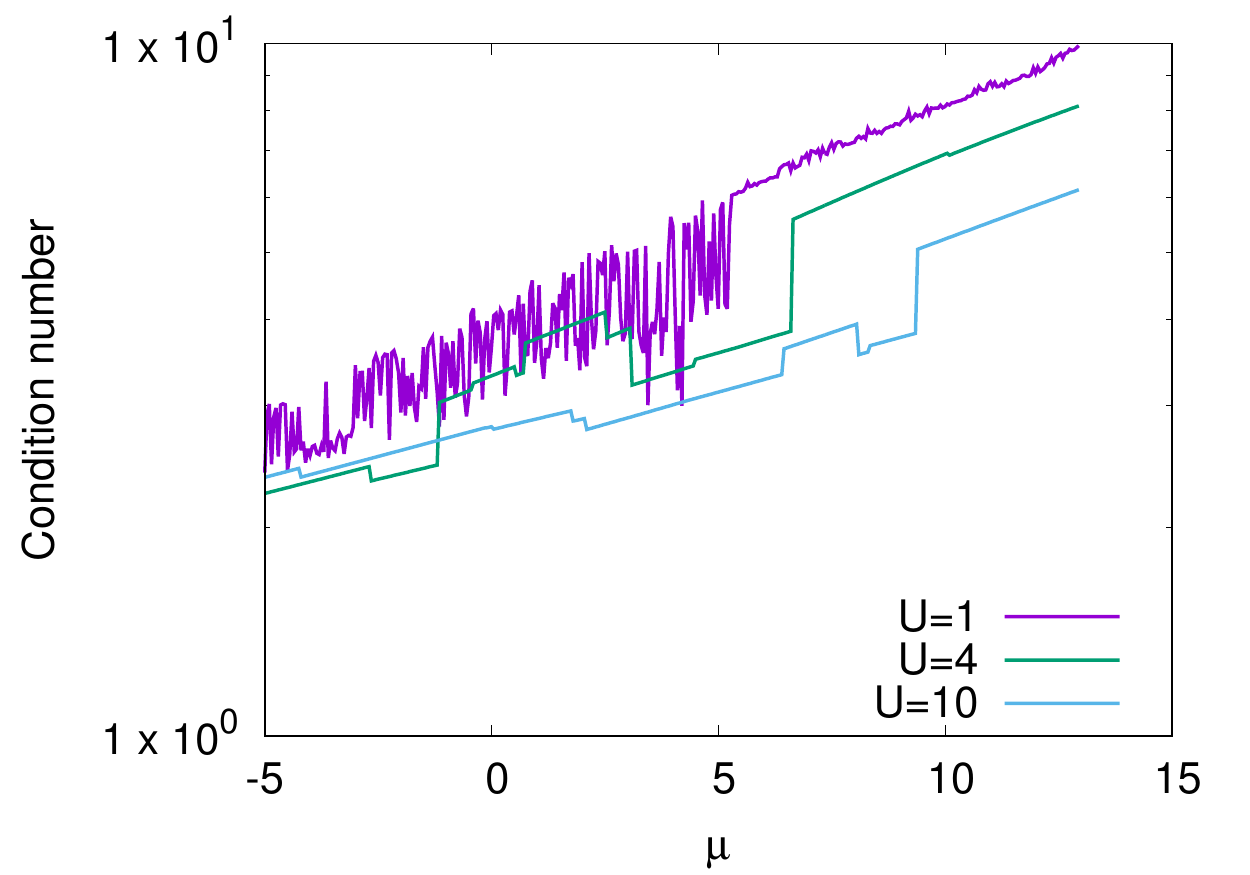} 
\caption{Condition number of the linear system \eqref{eq:linearsys} that is used to solve for the averages as a function of $\mu$ for $\beta = 0.1$.}
\label{fig:condbeta01} 
\end{figure}

\begin{figure}
\includegraphics[width=0.5\textwidth]{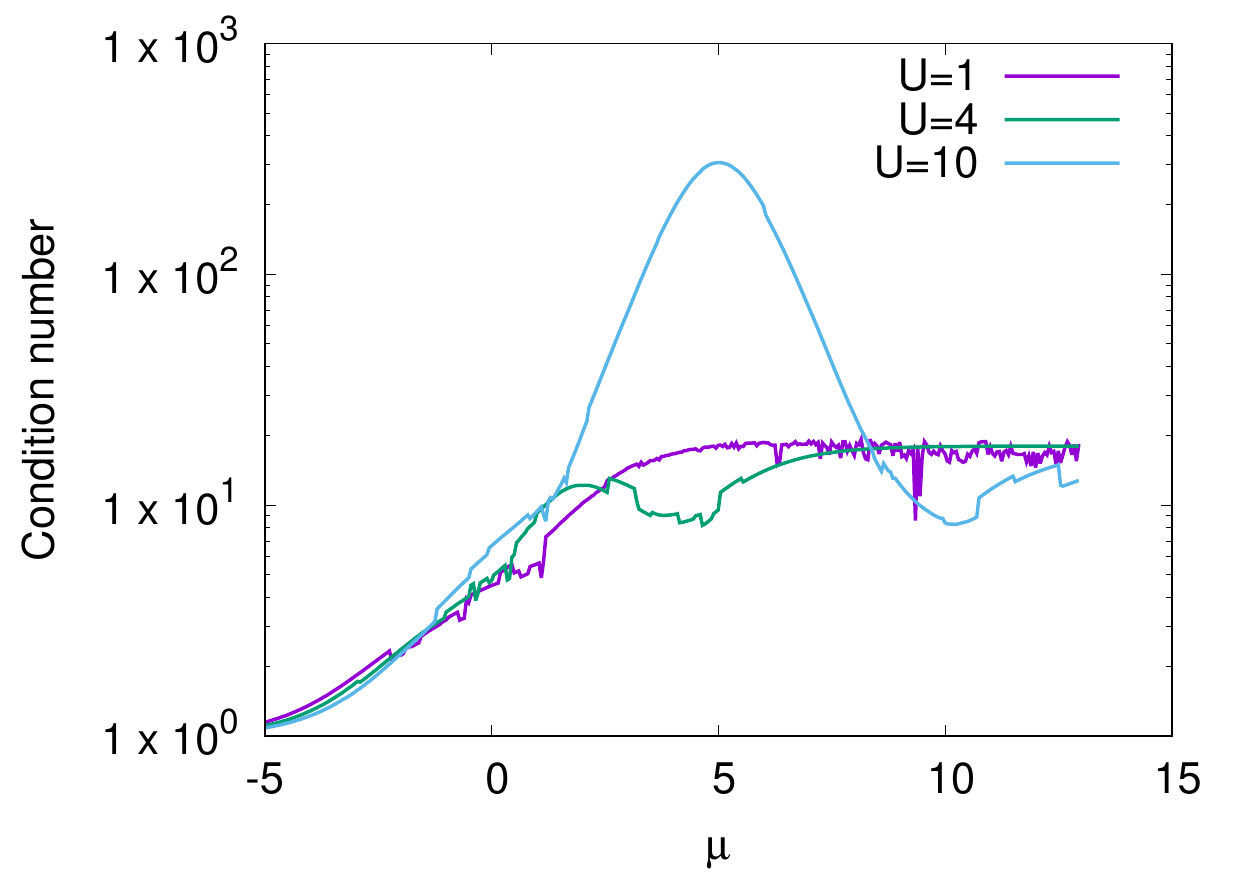}
\caption{Condition number of the linear system \eqref{eq:linearsys} that is used to solve for the averages as a function of $\mu$ for $\beta = 1$.}
\label{fig:condbeta1} 
\end{figure}

As stated previously, it is not guaranteed that the linear system we need to solve is numerically well-behaved.
Consequently in this method it is crucial to always check the condition number of the linear system for the desired set of free parameters $t$, $U$, $\beta$, $\mu$, in order to estimate the numerical error induced in the solution.
In fact let us assume that we need to solve a generic linear system in presence of numerical errors.
\begin{equation}
A(x+\delta x)=(b+\delta b) .
\end{equation}
It is then possible to show that\cite{trefethen1997numerical}
\begin{equation}
\frac{\Vert \delta x \Vert}{\Vert x \Vert} <  \text{cond}[A] \frac{\Vert \delta b \Vert}{\Vert b  \Vert} .
\end{equation} 
Usually the relative error induced in the inhomogeneous term is of the order of the numerical precision used on the machine $\epsilon_{mach}$. 
Consequently if the condition number $\text{cond}[A]$ is greater than $1/\epsilon_{mach}$ the solution will suffer from more than $100\%$ relative error and in the case where $\text{cond}[A]=0$ the system is actually unsolvable.
In Figs.~\ref{fig:condbeta01} and \ref{fig:condbeta1} we report the condition number of the linear system obtained for a few different values of the interaction strength $U$, two different values of the inverse temperature  $\beta \leq1$ as a function of the chemical potential $\mu$. In these cases the linear system is numerically well defined for all of the interaction strengths considered and we can determine the averages reliably.

\subsection{Condition number -- Low temperatures}

\begin{figure}[]
\includegraphics[width=0.5\textwidth]{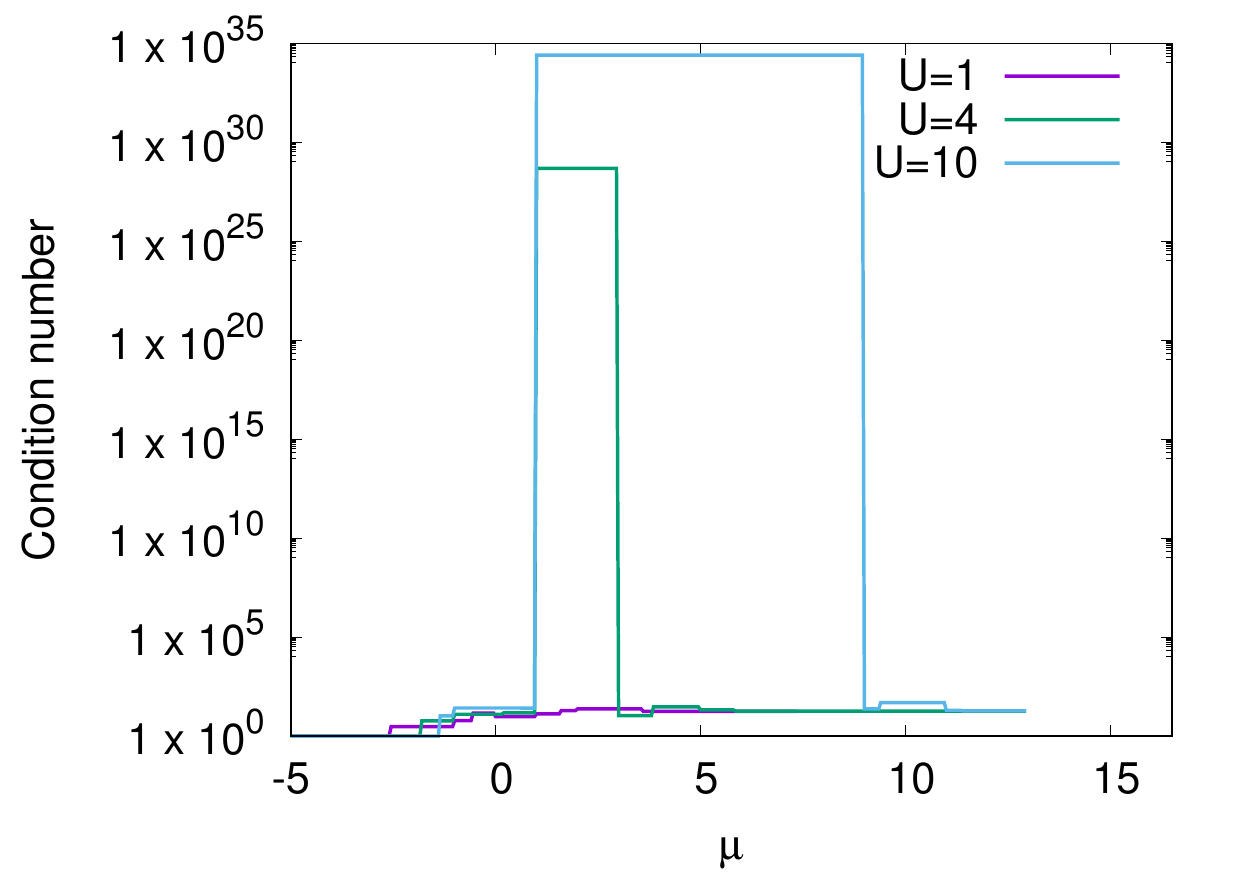} 
\caption{Condition number of the linear system \eqref{eq:linearsys} that is used to solve for the averages as a function of $\mu$ for $T = 0$.}
\label{fig:conmu0}
\end{figure}

\begin{figure}[]
\includegraphics[width=0.5\textwidth]{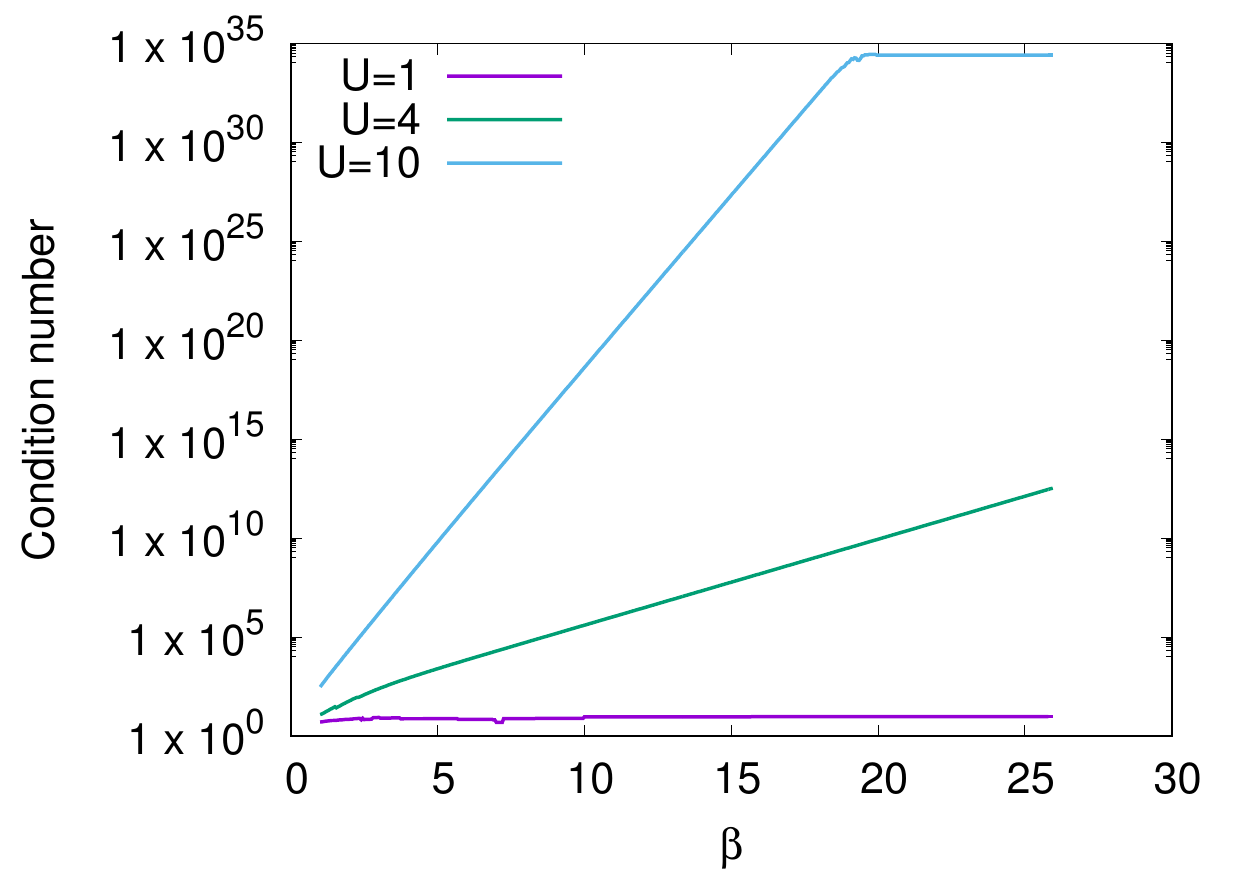} 
\caption{Condition number of the linear system \eqref{eq:linearsys} that is used to solve for the averages as a function of $\beta$ for $\mu = U/2$.}
\label{fig:conlow}
\end{figure}

Now let us focus on the  the condition number of the linear system in Eq.~\eqref{eq:linearsys} for  $\beta>1$.
In Fig.~\ref{fig:conmu0} we provide the result for zero temperature $\beta \to \infty$.
We find that the linear system is well defined for all chemical potentials for $U\in[0,2]$. 
For $U > 2$  we notice that the linear system becomes numerically ill-defined for chemical potentials in the range of $\mu \in [t,U-t]$.
We therefore consider the condition number for the system exactly at half-filling $\mu=U/2$ as a function of inverse temperature $\beta$, in order to understand how the linear system approaches its singular form in the limit $\beta \to \infty$. This result is shown in Fig.~\ref{fig:conlow}.
It is clearly seen that as $\beta$ increases the linear system display an asymptotic singular behavior and the divergence rate depends strongly on the interaction strength $U$. On the other hand the system is always solvable for all finite temperatures, provided that the precision used to represent the number is lower than the inverse of  the condition number of the linear system. 

Upon increasing the chemical potential there is no problem upon entering the 2-electron sector at $\mu_{12} = E_{GS}(2) - E_{GS}(1) = t - J$ where $J \approx 4t^2 /U $ when $t \ll U$. Here $E_{GS}(N)$ is the ground state energy for $N$ electrons in the absence of a chemical potential. The system becomes ill-behaved when $\mu$ is further increased to go above $\mu_C = E_{\text{triplet}}(2) - E_{GS}(1) = t$, i.e., when the free energy of the $N=2$ triplets go below that of the lowest energy state at $N=1$. For $\mu_{12} \leq \mu \leq \mu_C$ the zero temperature limit is well-defined and the correct ground state Green's function is obtained.
We interpret the problem with our scheme when $\mu > \mu_C$ that in this case the formalism has great difficulty to determine which of the 2-particle states that is the ground state.
One way to fix this problem is to add a term to the linear system that measures the total spin of the state. Solving the corresponding minimization problem we do obtain the correct Green's function by demanding that the $N=2$ ground state is a singlet.

\subsection{Observables and exact diagonalization}

With a quadruple precision calculation with  $U=10$ the lowest temperature we can determine the solution with $10\%$ of relative error is $T=0.055 t$ which corresponds to $\beta=18$. By using the method previously developed we compute the average site occupation $\bar{n} = \langle \hat{N}_{\text{total}} \rangle/4$ and double occupation $\langle \hat{n}_{\downarrow i} \hat{n}_{\uparrow i}\rangle$ as a function of the chemical potential and compare it with the one obtained from exact diagonalization.
\begin{figure}[]
\includegraphics[width=0.5\textwidth]{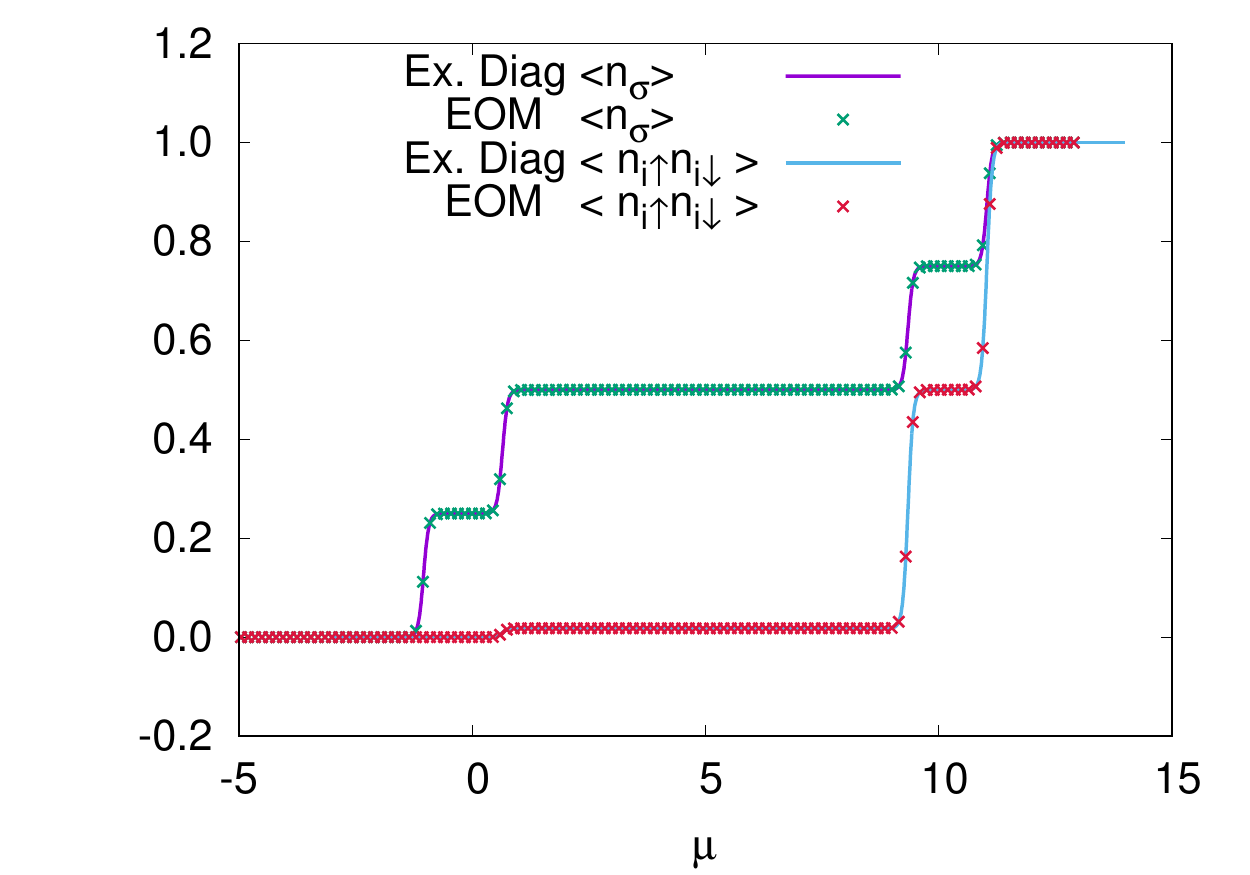} 
\caption{The occupancy and double occupancy of the system as a function of chemical potential $\mu$ using the proposed method (EOM), and compared with exact diagonalization (Ex. Diag) for $U=10$.}
\label{fig:occU10b18}
\end{figure}
Representative results are shown in Fig.~\ref{fig:occU10b18} and the result obtained with the proposed scheme are clearly in good agreement with the exact solution.

\begin{figure}[]
\includegraphics[width=0.5\textwidth]{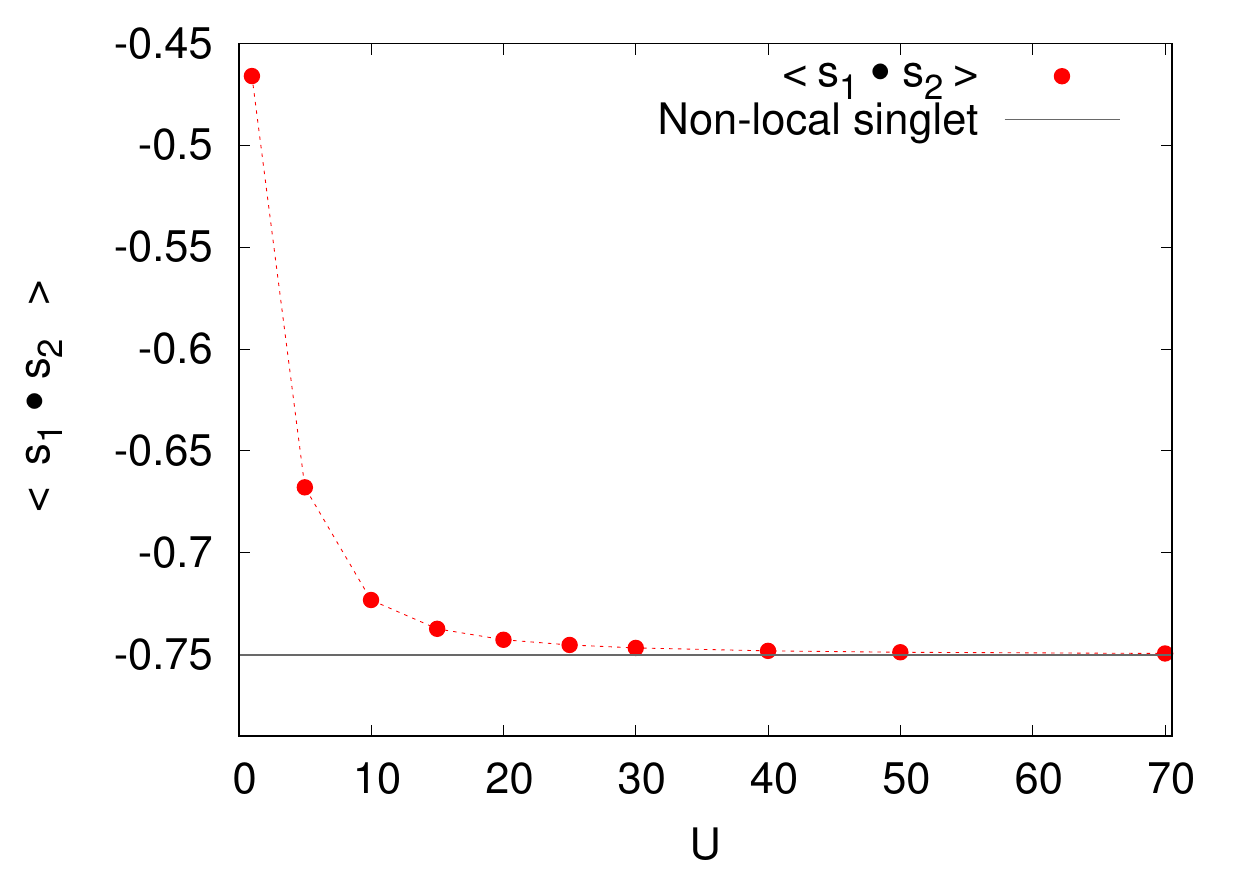} 
\caption{Spin-spin correlation function $\langle {\bf S}_1 \cdot{\bf S}_2 \rangle$ as a function of Hubbard interaction $U$ at half filling obtained with the proposed method. The reference at $-3/4$ is the correlation of the non-local singlet.}
\label{fig:spsp}
\end{figure}

At a chemical potential $\mu=t-\epsilon$, with $\epsilon$ a small constant number, the linear system is well behaved as discussed above and we obtain a  density $\la \hat{n}_{i\sigma} \ra=0.5$, which corresponds to half-filling. Consequently we can address some half-filling properties at zero temperature. In particular we have studied how the spin-spin correlation $\langle {\bf S}_1 \cdot{\bf S}_2 \rangle$ behaves as a function of the interaction strength $U$. The result is shown in Fig.~\ref{fig:spsp}. For large values of $U$ the correlations are those of the non-local singlet as expected.

\section{Conclusions and Outlook}
\label{sec:conclusions}

In this work we have mainly focused on one aspect of the EOM for Green's functions, namely the determination of the averages that goes into the Green's function.  For our scheme to work it is necessary to write down a relation like that in Eq.~\eqref{av:anti}. This is usually unproblematic.
It is also crucial to note that we do not consider the whole Green's function matrix $\dla A_i | A_j^\dagger \dra_z$ for all $i$ and $j$, but only $\dla A_i | B^\dagger \dra_z$ for all $i$ and one or a few fixed $B$'s. This reduces the number of unknowns that must be determined, and also simplifies the algebra considerably.
The main appeal of this scheme is that there is a linear system to solve for the unknown averages, which is a simple matter when the system is well-conditioned. For high temperatures this is always the case, but for low temperatures it is not always so.
We have identified one cause of this problem that is clear when increasing the chemical potential $\mu$. Upon entering a new charge sector $Q$ there is no problem: all excitation energies from the new ground state to the lowest energy charged states (in sectors with charge $Q \pm 1$) are positive. Also all excitation energies between the lowest energy states with $Q \pm 1$ and excited states in the charge $Q$ sector are positive.
Upon increasing the chemical potential further also some additional excitation energies just mentioned goes negative. The linear system then becomes close to singular and the system does not know which of the charge $Q$ states that is the lowest energy one. On the other hand we know that nothing special should happen at this point: the lowest Free energy state do not change, so all averages should remain essentially unchanged when crossing this value of the chemical potential, even at zero temperature. Indeed, when substituting the averages obtained below this critical value of the chemical potential into the linear system we find that it is still a solution also above this value at zero temperature. This is another way to solve the ill-conditioning problem in the zero temperature limit. Another possibility that we discussed is to use some additional quantum numbers, such as total spin, to restrict the ground state.

In this work we have assumed that the EOM close, i.e., that the exact $K$ matrix is known. This allows us to focus on the determination of the averages. On the other hand, in realistic extended many-body system this is never the case and a procedure for truncation and including the physics of the neglected operators is necessary. A detailed study of this problem, and the connection to the scheme proposed in this paper, is planned for future work.

\acknowledgments

Funding from the Knut and Alice Wallenberg Foundation and the Swedish research council Vetenskapsr{\aa}det is gratefully acknowledged.

\bibliography{eom-refs}

\begin{widetext}

\newpage

\appendix

\section{Operator basis}
\label{app:A}

The operator basis used in the main text is given in the following equation.
For notational convenience we here use $c$ for spin-up and $d$ for spin-down, i.e., we write $c_{1\ua} =c_{1} $, $c_{1\da}=d_{1} $, etc.
\begin{equation}
\label{eq:operatorbasiscomplete}
\begin{split}
A_1=&(1-d^+_1d_1)c_1\\
A_2=&d^+_1d_1c_1\\
A_3=&(1-d^+_2d_2)c_2\\
A_4=&d^+_2d_2c_2\\
A_5=&-d^+_1d_2c_1\\
A_6=&-d^+_1d_1c_1\\
A_7=&-d^+_2d_1c_1\\
A_8=&-d^+_1d_2c_2\\
A_9=&-d^+_2d_2c_1\\
A_{10}=&-d^+_2d_1c_2\\
A_{11}=&c^+_2d_1^+c_1c_2d_2\\
A_{12}=&d_1^+d_2^+c_2d_1d_2\\
A_{13}=&-(c_2^+d_2^+c_1c_2d_1)\\
A_{14}=&c_1^+d_1^+c_1c_2d_2\\
A_{15}=&d_1^+d_2^+c_1d_1d_2\\
A_{16}=&-(c_1^+d_2^+c_1c_2d_1)\\
A_{17}=&c_2^+d_1^+c_1c_2d_1\\
A_{18}=&c_1^+d_2^+c_1c_2d_2\\
A_{19}=&c_1^+d_1^+c_1c_2d_1\\
A_{20}=&c_1^+d_2^+c_1c_2d_2\\
A_{21}=&-(c_2^+d_1^+d_2^+c_1c_2d_1d_2)\\
A_{22}=&-(c_1^+d_1^+d_2^+c_1c_2d_1d_2)
\end{split}
\end{equation}

\section{Matrix elements}
\label{app:B}
The non-zero matrix elements of $K$, $M$, and $v$ using the basis of Eq.~\eqref{eq:operatorbasiscomplete} are given in the following table.
\begin{table}[h]
\begin{tabular}{|l|l|l|l|l|l|l|l|l|}
\hline
$K_{1,3}=t$  & $K_{3,10}=t$  & $K_{7,2}=t$  & $K_{9,15}=U$   & $K_{13,18}=-t$ & $K_{17,17}=U$  & $K_{21,22}=-t$ & $M_{9,2}=-1$   & $v_{1}=1$ \\ \hline
$K_{1,4}=t$  & $K_{4,8}=t$   & $K_{7,9}=t$  & $K_{10,6}=-t$  & $K_{14,11}=-t$ & $K_{18,20}=-t$ & $K_{21,21}=U$  & $M_{11,7}=1$   &         \\ \hline
$K_{1,5}=t$  & $K_{4,9}=-t$  & $K_{7,10}=t$ & $K_{10,7}=t$   & $K_{14,19}=t$  & $K_{18,11}=-t$ & $K_{22,21}=-t$ & $M_{13,5}=-1$  &         \\ \hline
$K_{1,6}=t$  & $K_{4,10}=-t$ & $K_{7,7}=U$  & $K_{10,4}=-t$  & $K_{14,20}=-t$ & $K_{18,13}=-t$ & $K_{22,22}=U$  & $M_{14,8}=1$   &         \\ \hline
$K_{1,7}=-t$ & $K_{4,4}=U$   & $K_{7,13}=U$ & $K_{10,16}=U$  & $K_{15,12}=t$  & $K_{18,21}=U$  & $M_{1,1}=-1$   & $M_{15,9}=1$   &         \\ \hline
$K_{2,5}=-t$ & $K_{5,2}=-t$  & $K_{8,5}=t$  & $K_{11,14}=-t$ & $K_{15,15}=U$  & $K_{19,14}=t$  & $M_{1,2}=-1$   & $M_{15,10}=-1$ &         \\ \hline
$K_{2,6}=-t$ & $K_{5,8}=t$   & $K_{8,6}=t$  & $K_{11,17}=t$  & $K_{16,13}=-t$ & $K_{19,16}=t$  & $M_{2,1}=1$    & $M_{16,10}=-1$ &         \\ \hline
$K_{2,7}=t$  & $K_{5,9}=-t$  & $K_{8,4}=t$  & $K_{11,18}=-t$ & $K_{16,19}=t$  & $K_{19,17}=-t$ & $M_{2,2}=1$    & $M_{17,9}=1$   &         \\ \hline
$K_{2,2}=U$  & $K_{5,11}=U$  & $K_{8,8}=u$  & $K_{11,11}=U$  & $K_{16,20}=-t$ & $K_{19,22}=U$  & $M_{5,3}=-1$   & $M_{18,2}=-1$  &         \\ \hline
$K_{3,1}=t$  & $K_{6,2}=-t$  & $K_{8,14}=U$ & $K_{12,15}=t$  & $K_{16,16}=U$  & $K_{20,14}=-t$ & $M_{5,4}=-1$   & $M_{19,6}=1$   &         \\ \hline
$K_{3,2}=t$  & $K_{6,8}=t$   & $K_{9,5}=-t$ & $K_{12,12}=U$  & $K_{17,19}=-t$ & $K_{20,16}=-t$ & $M_{7,3}=-1$   & $M_{20,4}=-1$  &         \\ \hline
$K_{3,8}=-t$ & $K_{6,10}=-t$ & $K_{9,7}=t$  & $K_{13,16}=-t$ & $K_{17,11}=t$  & $K_{20,18}=-t$ & $M_{7,4}=-1$   & $M_{21,18}=1$  &         \\ \hline
$K_{3,9}=t$  & $K_{6,12}=U$  & $K_{9,4}=-t$ & $K_{13,17}=t$  & $K_{17,13}=t$  & $K_{20,20}=U$  & $M_{9,1}=-1$   & $M_{22,12}=1$  &         \\ \hline
\end{tabular}
\end{table}

\end{widetext}

\end{document}